\documentclass[a4paper,11pt]{article}
\usepackage{pos}
\usepackage{braket}
\usepackage{comment}

\title{Determination of the CP restoration temperature at $\theta=\pi$ in 4D SU(2) Yang-Mills theory through simulations at imaginary $\theta$}
\ShortTitle{Determination of the CP restoration temperature at $\theta=\pi$ in 4D SU(2) YM theory}

\author*[a,b]{Mitsuaki Hirasawa}
\author[c]{Kohta Hatakeyama}
\author[d,e]{Masazumi Honda}
\author[d,e]{Akira Matsumoto}
\author[f, g]{Jun Nishimura}
\author[h]{Atis Yosprakob}

\affiliation[a]{Dipartimento di Fisica, Università di Milano-Bicocca, \\
Piazza della Scienza 3, I-20126, Milano, Italy}
\affiliation[b]{Sezione di Milano Bicocca, Istituto Nazionale di Fisica Nucleare,\\ Piazza della Scienza, 3, I-20126 Milano, Italy}
\affiliation[c]{Graduate School of Science and Technology, Hirosaki University,\\ Bunkyo-cho 3, Hirosaki, Aomori 036-8561, Japan}
\affiliation[d]{RIKEN iTHEMS, 2-1 Hirosawa, Wako, Saitama 351-0198, Japan}
\affiliation[e]{Yukawa Institute for Theoretical Physics, Kyoto University,\\ Sakyo-ku, Kyoto 606-8502, Japan}
\affiliation[f]{KEK Theory Center, High Energy Accelerator Research Organization,\\ 1-1 Oho, Tsukuba, Ibaraki 305-0801, Japan}
\affiliation[g]{Graduate Institute for Advanced Studies, SOKENDAI,\\ 1-1 Oho, Tsukuba, Ibaraki 305-0801, Japan}
\affiliation[h]{Department of Physics, Niigata University, Niigata 950-2181, Japan.}

\emailAdd{Mitsuaki.Hirasawa@mib.infn.it}
\emailAdd{kohta.hatakeyama@gauge.scphys.kyoto-u.ac.jp}
\emailAdd{masazumi.honda@riken.jp}
\emailAdd{akira.matsumoto@yukawa.kyoto-u.ac.jp}
\emailAdd{jnishi@post.kek.jp}
\emailAdd{ayosp@phys.sc.niigata-u.ac.jp}

\abstract{The 't Hooft anomaly matching condition provides constraints on the phase structure at $\theta=\pi$ in 4D SU($N$) Yang-Mills theory. 
In particular, 
assuming that the theory is confined and the CP symmetry is spontaneously broken at low temperature,
it cannot be restored below the deconfining temperature at $\theta=\pi$. 
Here we investigate the CP restoration at $\theta=\pi$ in the 4D SU(2) case and provide numerical evidence that the CP restoration occurs at a temperature higher than the deconfining temperature unlike the known results in the large-$N$ limit, where the CP restoration occurs precisely at the deconfining temperature. The severe sign problem at $\theta=\pi$ is avoided by focusing on the tail of the topological charge distribution at $\theta=0$, which can be probed by performing simulations at imaginary $\theta$. 
By analytic continuation with respect to $\theta$, we obtain the topological charge at real $\theta$.}

\FullConference{The 40th International Symposium on Lattice Field Theory (Lattice 2023)\\
July 31st - August 4th, 2023\\
Fermi National Accelerator Laboratory\\}

\begin{document}
\begin{flushright}
KEK-TH-2594, RIKEN-iTHEMS-Report-23, YITP-24-03
\end{flushright}
\maketitle

\section{Introduction}
Recently the phase structure of the gauge theory with a $\theta$ term has been studied both analytically and numerically \cite{Gaiotto:2017yup, Kitano:2021jho, Chen:2020syd,Cordova:2019bsd}.
Especially, the authors of Ref.~\cite{Gaiotto:2017yup} 
discussed the 't Hooft anomaly matching condition \cite{tHooft:1979rat} involving the generalized symmetry \cite{Gaiotto:2014kfa}, 
and found that there is a constraint on a relation between the spontaneous symmetry breaking (SSB) of the CP and $Z_N$ center symmetries at $\theta = \pi$.

It is known that there is a CP broken confined phase in the large-$N$ Yang-Mills theory (YM), and the critical temperatures of the SSB of CP and $Z_N$ coincide \cite{Witten:1980sp,Witten:1998uka}.
On the other hand, in the small $N$ case, the phase structure is not clear.
Recent lattice studies suggest that the theory is in the CP broken phase at $T=0$ for $\theta=\pi$ \cite{Kitano:2021jho}. 
By combining it with the well-known fact that the theory is in the CP preserving deconfined phase at high temperature \cite{Gross:1980br,Weiss:1980rj}, the CP restoration occurs at finite temperature.
Using the anomaly matching condition one can say that the CP restoration temperature $T_{\rm CP}$ is equal to or higher than the deconfining temperature $T_{\rm dec}$ at $\theta=\pi$.
If the latter possibility $T_{\rm CP} > T_{\rm dec}$ is true, then it implies that the pure YM 
has a deconfined phase with spontaneously broken CP symmetry in the intermediate temperature regime.   
The previous study \cite{Chen:2020syd} argued that such an exotic phase exists in the SU(2) supersymmetric YM with a gaugino mass deformation which flows to the pure SU(2) YM in IR.

In this work, we perform the first principles calculation focusing on the fact that the order parameter of the CP symmetry at finite $\theta$ is related to the topological charge distribution at $\theta=0$.
In particular, the asymptotic behavior of the distribution distinguishes the CP symmetric and broken phases.
Since it is difficult to determine the asymptotic behavior by simulations at $\theta=0$, we use the imaginary $\theta$ parameter.
The imaginary and real $\theta$ dependences of the topological charge are related by an analytic continuation of $\theta$.
Similar methods were applied to the ${\rm CP}^9$ model and the Schwinger model in Refs.~\cite{Azcoiti:2003qe, Azcoiti:2017mxl}.
In particular, we discuss how we can determine the critical temperature using the analytic continuation.

\section{The SU($N$) gauge theory with the $\theta$ term}
The SU($N$) gauge theory with a $\theta$ term is defined by the following partition function:
\begin{equation}
    Z=\int \mathcal{D}A_\mu e^{-S_{\rm g} + i\theta Q},
\end{equation}
where $Q$ is the topological charge defined by
\begin{equation}
    Q=\frac{1}{32\pi^2}\int d^4x \epsilon_{\mu\nu\rho\sigma} {\rm Tr}\left[ F_{\mu\nu} F_{\rho\sigma} \right].
\end{equation}
The topological charge is an integer value $Q\in\mathbb{Z}$ on a compact manifold like a 4D torus.
As a consequence of this fact, this model has the CP symmetry not only at $\theta=0$ but also at $\theta=\pi$.
Under the CP transformation, the sign of topological charge flips while the action $S_{\rm g}$ is invariant.
Therefore the CP transformation corresponds to a sign flip of the theta ($\theta \to -\theta$).

The behaviour of topological charge expectation value $\braket{Q}_{\theta}$ has the following equality around $\theta = \pi$:
\begin{equation}
\begin{split} 
    \braket{Q}_{\theta = \pi-\epsilon} &= - \braket{Q}_{\theta = -(\pi-\epsilon)}\\
    &=-\braket{Q}_{\theta = \pi+\epsilon},
\end{split}
\end{equation}
where we use the CP symmetry and the $2\pi$ periodicity: $\theta = \theta + 2\pi n$ for the first and second equalities, respectively.
Using this equality, we define an order parameter for the SSB of the CP symmetry as
\begin{equation}
    \lim_{\epsilon\to0} \lim_{V_{\rm s}\to\infty} \frac{\Braket{Q}_{\theta=\pi-\epsilon}}{V_{\rm s}}.
\end{equation}
This order parameter is nonzero in the CP broken phase, while it is zero in the CP restored phase.
In other words, $\Braket{Q}_\theta$ changes discontinuously near $\theta = \pi$ in the CP broken phase, and it changes continuously in the CP restored phase.
Due to the sign problem, it is difficult to study this directly by Monte Carlo simulations.

Here we focus on the relation between $\Braket{Q}_\theta$ at non-zero $\theta$ and the topological charge distribution at $\theta=0$.
The partition function at $\theta$ can be written as
\begin{equation}
    Z_\theta = \int dU e^{-S_g+i\theta Q} = Z_0\int dq e^{i\theta q}\rho(q),
\end{equation}
where $\rho(q)$ is the topological charge distribution at $\theta=0$ defined by
\begin{equation}
    \rho(q) = \frac{1}{Z_0}\int dU \delta(q-Q)e^{-S_g}. 
\end{equation}
Using this partition function, we can write the expectation value of the topological charge at arbitrary $\theta$ as
\begin{equation}
    \Braket{Q}_\theta = -i\frac{\partial}{\partial\theta}\log{Z_\theta} = \frac{\int dq q e^{i\theta q}\rho(q)}{\int dq e^{i\theta q}\rho(q)}.
\label{eq:Q_theta}
\end{equation}

\section{$\theta$ dependence of $\Braket{Q}_\theta$ in two models}
\begin{figure}
    \centering
    \includegraphics[width=0.45\hsize]{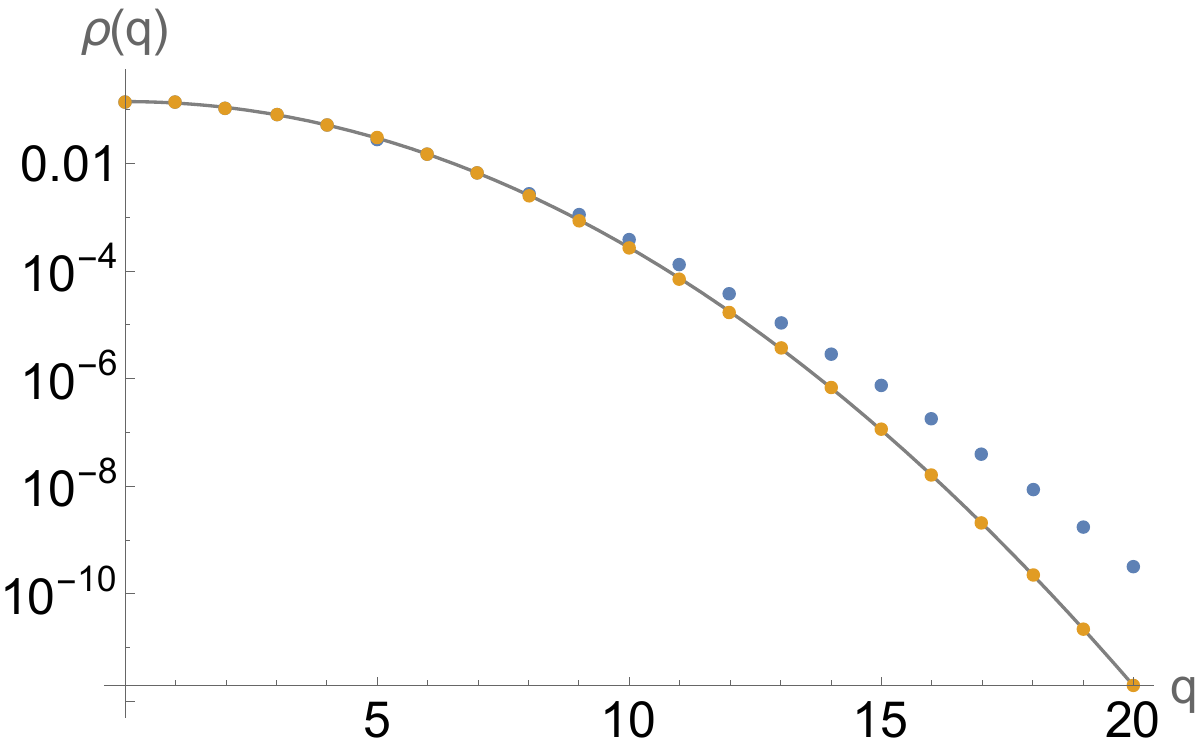}
    \caption{Topological charge distribution in the logarithmic scale for the instanton gas model (blue) and the Gaussian model (orange). The solid line represents a Gaussian behavior $\propto\exp\left(-q^2/2\chi_0V\right)$.
    We can see small discrepancies only at large $q$.}
    \label{fig:rho_q}
\end{figure}

In this section we discuss the $\theta$ dependence of $\Braket{Q}_\theta$ in two simplified models in which the behaviours of $\Braket{Q}_\theta$ are different.
One is the instanton gas model which is an effective model of the Yang-Mills theories at sufficiently high temperature where the CP is restored.
The other is the Gaussian model which is an effective model of the large-$N$ Yang-Mills theory at low temperature where the CP is broken.
The asymptotic behaviour of the topological charge distribution at large $|q|$ is given as
\begin{equation}
    \rho(q) \sim
    \left\{
        \begin{array}{ll}
        \exp{(-q \log{q})} &:\ {\rm instanton\ gas\ model},\\
        \exp{(-\frac{q^2}{2\chi_0V})} &:\ {\rm Gaussian\ model }\ ({\rm low}\ T),
        \end{array}
    \right.
\label{eq:q_dists}
\end{equation}
where $\chi_0$ is the topological susceptibility at $\theta = 0$.
Therefore, the CP restoration at $\theta=\pi$ is related to the asymptotic behaviour of $\rho(q)$.
In Fig.\ref{fig:rho_q}, we plot $\rho(q)$ in the logarithmic scale against $q$ for the two cases.
We can see a tiny difference in the tail of the distributions.
Using the Eq.~(\ref{eq:Q_theta}) and Eq.~(\ref{eq:q_dists}), we obtain the $\theta$ dependence of $\Braket{Q}_\theta$ in these two models as
\begin{equation}
    \frac{\braket{Q}_\theta}{\chi_0 V} = 
    \left\{
        \begin{array}{ll}
        i \sin \theta &:\ {\rm instanton\ gas\ model},\\
        i \theta &:\ {\rm Gaussian\ model }\ ({\rm low}\ T).
        \end{array}
    \right.
\end{equation}

Since the probability of obtaining gauge configurations with a large topological charge is highly suppressed, it is not feasible to observe the asymptotic behaviour of the $\rho(q)$ directly by performing lattice simulations at $\theta=0$.
Therefore, we focus on the fact that the asymptotic behaviour is enhanced at imaginary $\theta$ due to the weight $e^{\tilde{\theta}Q}$ where $\tilde{\theta} = i\theta \in \mathbb{R}$:
\begin{equation}
    \Braket{Q}_{\tilde{\theta}} = \frac{\int dq qe^{\tilde{\theta}q}\rho(q)}{\int dq e^{\tilde{\theta}q}\rho(q)} .
\end{equation}
In the above two models, by performing the analytic continuation $\tilde{\theta} = i\theta$, we obtain the imaginary theta $\tilde{\theta}$ dependence of $\Braket{Q}_{\tilde{\theta}}$ as
\begin{equation}
    \frac{\braket{Q}_{\tilde{\theta}}}{\chi_0 V} = 
    \left\{
        \begin{array}{ll}
        \sinh \tilde{\theta} &:\ {\rm instanton\ gas\ model,}\\
        \tilde{\theta} &:\ {\rm Gaussian\ model }\ ({\rm low}\ T).
        \end{array}
    \right.
\label{eq:q_im_theta}
\end{equation}
As we can see in this equation, at imaginary theta $\tilde{\theta} = i \theta$, $\Braket{Q}_{\tilde{\theta}}$ is expected to approach the hyperbolic sine function of $\tilde{\theta}$ at high temperature and the linear function of $\tilde{\theta}$ at low temperature also in the SU(2) case.
Therefore, it is possible to observe the difference of the asymptotic behaviour of the topological charge distribution by performing lattice simulations at $\tilde{\theta}$.

\section{Lattice setup}
In our simulations, we use the Wilson gauge action defined by
\begin{equation}
S_{\rm g}=-\frac{\beta}{2N}\sum_{n}\sum_{\mu\neq\nu}\mathrm{Tr}\left[ P_{n}^{\mu\nu} (U) \right],
\end{equation}
where $\beta = 2/g_0^2$ and the plaquette is defined by
\begin{equation}
P_{n}^{\mu\nu}(U)=U_{n,\mu}U_{n+\mu,\nu}U_{n+\nu,\mu}^{\dagger}U_{n,\nu}^{\dagger}.
\end{equation}
In this work, we determine the temperature using the $\beta$ function as in Ref.~\cite{Engels:1994xj}.
For the lattice implementation of the topological charge, we use the clover-leaf definition \cite{DiVecchia:1981aev}
\begin{equation}
Q[U]=-\frac{1}{32\pi^{2}}\sum_{n}\frac{1}{2^{4}}\sum_{\mu,\nu,\rho,\sigma=\pm1}^{\pm4}\tilde{\epsilon}_{\mu\nu\rho\sigma}\mathrm{Tr}\left[P_{n}^{\mu\nu}P_{n}^{\rho\sigma}\right],
\end{equation}
where $\tilde{\epsilon}_{\mu\nu\rho\sigma}$ is an anti-symmetric tensor defined by 
\begin{equation}
\tilde{\epsilon}_{1234}=-\tilde{\epsilon}_{2134}=-\tilde{\epsilon}_{-1234}=\cdots = 1.
\end{equation}
In order to remove the UV fluctuations from the topological charge, we use the stout smearing method \cite{Morningstar:2003gk}.
Let $\tilde{U}$ denote the smeared (or fat) link.
For the lattice simulations, we use the following action:
\begin{equation}
    S = S_{\rm g} + \tilde{\theta}~Q[\tilde{U}].
\label{lattice_action}
\end{equation}
In our simulations, we set the smearing step size as 0.09 and the number of smearing step as 40, for which we observe the comb-like structure in the topological charge distribution.

To update the gauge configurations, we use the Hybrid Monte Carlo method.
The drift force comes not only from the gauge action but also from the topological term.
Since we use the smeared link for the topological charge in Eq.~(\ref{lattice_action}), we need to calculate the drift force for the original link by reversing the smearing steps.
For the detail, see Refs.~\cite{Matsumoto:2021zjf, Matsumoto:2023vbf}.

\section{Results}
Our analysis consists of three steps.
First, we need to make an infinite volume extrapolation for $\Braket{Q}_{\tilde{\theta}}$, which is important in studying the behaviour near the first-order phase transition.
Next, we make an infinite volume extrapolation for the topological charge susceptibility $\chi_0$.
Then, we fit the extrapolated values of $\Braket{Q}_{\tilde{\theta}}$ to polynomial and hyperbolic sine series, where the coefficients of the leading terms in both series are fixed by $\chi_0$.
By performing the analytic continuation of the fitting functions, we observe the CP breaking/restoration.
We fix the lattice size for the temporal direction as $L_{\rm t}=5$ in this work.

\subsection{Infinite volume extrapolation for $\Braket{Q}_{\tilde{\theta}}/V_{\rm s}$}
\begin{figure}
    \centering
    \includegraphics[width=0.45\hsize]{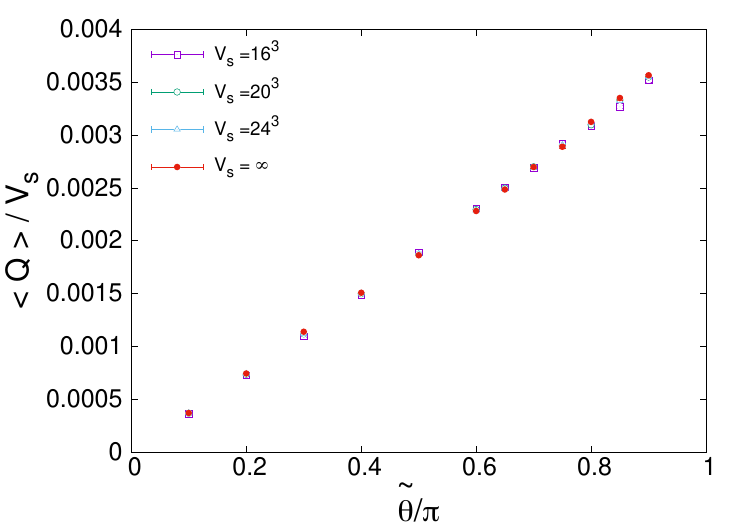}
    \includegraphics[width=0.45\hsize]{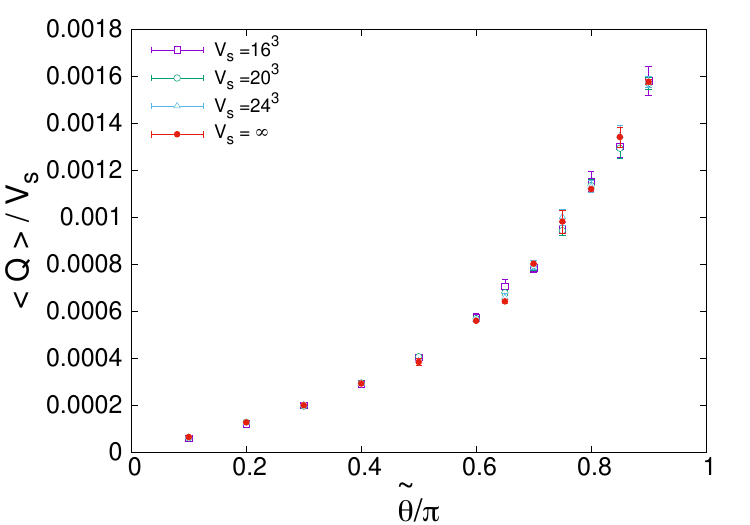}
    \caption{$\Braket{Q}/V_{\rm s}$ is plotted against $\tilde{\theta}/\pi$ for $T=0.9T_{\rm c}$ (Left) and $T=1.2T_{\rm c}$ (Right). We cannot see any significant volume dependence,}
    \label{fig:infinite_V}
\end{figure}

We use the data obtained with the spatial lattice size $L_{\rm s} = 16, 20, 24$ for the infinite volume extrapolation.
At each $\tilde{\theta}$, we fit the data points to the linear function $f(V_{\rm s}) = c + b/V_{\rm s}$, where $V_{\rm s} = L_{\rm s}^3$, $c$ and $b$ are the fitting parameters.
In Fig.\ref{fig:infinite_V}, we plot $\Braket{Q}_{\tilde{\theta}}/V_{\rm s}$ against $\tilde{\theta}$ at various values of the volume and the extrapolated value.
We can see from the plot that there is no significant volume dependence in $\Braket{Q}_{\tilde{\theta}}/V_{\rm s}$.

\subsection{Infinite volume extrapolation for $\chi_0$}
\begin{figure}
    \centering
    \includegraphics[width=0.45\hsize]{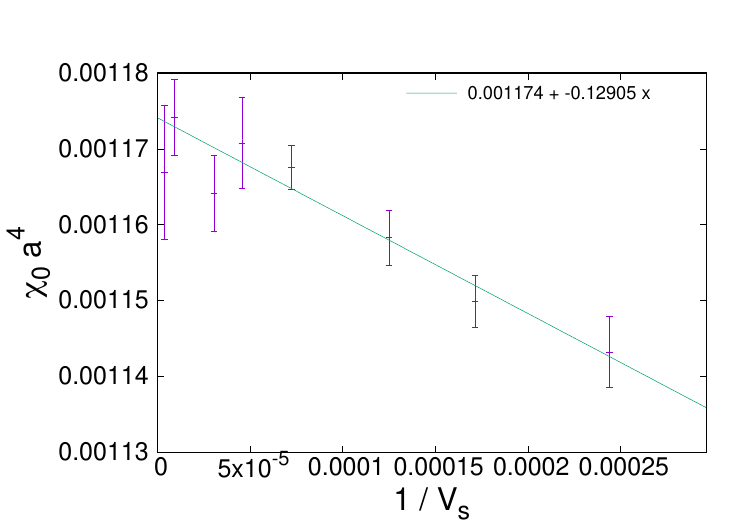}
    \includegraphics[width=0.45\hsize]{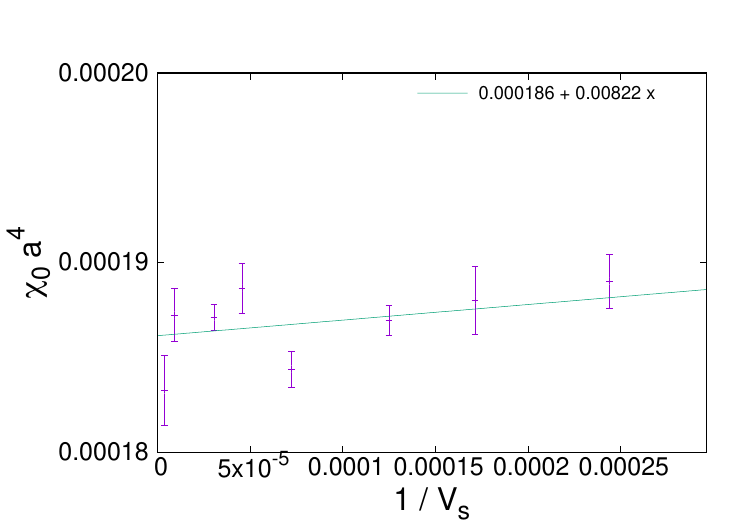}
    \caption{The quantity $\chi_0 a^4$, where $a$ is lattice spacing, is plotted against $1/V_{\rm s}$. The solid lines represent the fitting function $f(V_{\rm s}) = c+b/V_{\rm s}$ used for the infinite volume extrapolation.
    The extrapolated value of $\chi_0 a^4$ at $V_{\rm s} = \infty$ is 0.001174 and 0.000186 for $T=0.9T_{\rm c}$ and $T=1.2T_{\rm c}$, respectively. }
    \label{fig:chi0}
\end{figure}

We use the data points at various values of $L_{\rm s}$ from 16 to 64 in order to obtain the extrapolated values of $\chi_0$.
In Fig.\ref{fig:chi0}, we plot the $\chi_0$ against $1/V_{\rm s}$.
We fit the data points to the linear function $f(V_{\rm s}) = c + b/V_{\rm s}$, where $c$ and $b$ are the fitting parameters.

\subsection{Fitting}
Here we use the following two fitting functions motivated by the behaviour of $\Braket{Q}_{\tilde{\theta}}$ in the instanton gas model and the Gaussian model:
\begin{equation}
\begin{split}
    f(\tilde{\theta}) &= (\chi_0 - 2a_2 - 3a_3) \sinh{\tilde{\theta}} + a_2 \sinh{2\tilde{\theta}} + a_3 \sinh{3\tilde{\theta}}, \\
    g(\tilde{\theta}) &= \chi_0 \tilde{\theta} + b_3 \tilde{\theta}^3,
\end{split}
\end{equation}
where the coefficients of the leading terms are determined by $\left.\frac{\partial_{\theta} \Braket{Q}_\theta}{V}\right|_{\theta=0} = \chi_0$.
Since the topological charge is a CP odd quantity, only the odd powers of $\tilde{\theta}$ appear in $g(\tilde{\theta})$.

\begin{figure}
    \centering
    \includegraphics[width=0.45\hsize]{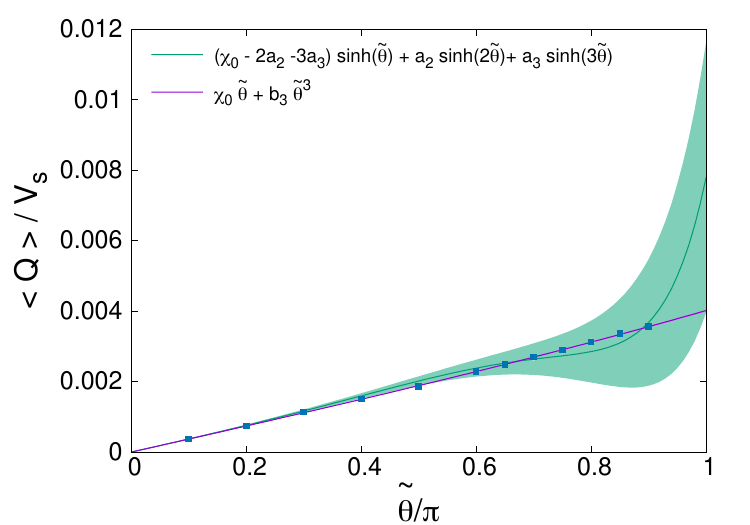}
    \includegraphics[width=0.45\hsize]{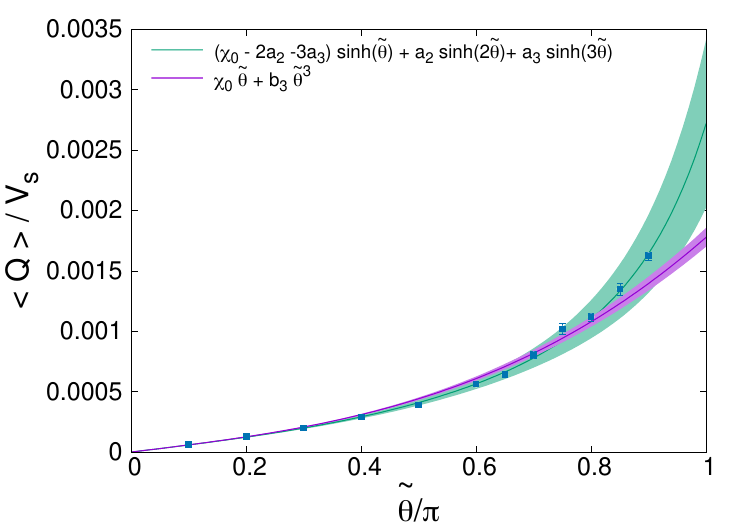}
    \caption{$\Braket{Q}_{\tilde{\theta}}/V_{\rm s}$ obtained by the infinite volume extrapolation is plotted against $\tilde{\theta}/\pi$ for $T=0.9T_{\rm c}$ (Left) and $T=1.2T_{\rm c}$ (Right). We fitted the data points to both the polynomial and hyperbolic sine series at each temperature.
    The values of the fitting parameters are given in Tab.~\ref{tab:fit_parm}.}
    \label{fig:fitting}
\end{figure}

In Fig.\ref{fig:fitting} (Left), we fit the data points to polynomial and hyperbolic sine series at $T = 0.9 T_{\rm c}$.
We can see that the fitting by polynomial works well, while the fitting by hyperbolic sine series has a relatively large discrepancy.
On the other hand, in Fig.\ref{fig:fitting} (Right), we fit the data points to polynomial and hyperbolic sine series at $T = 1.2 T_{\rm c}$.
We can see that the fitting by hyperbolic sine series works well, while the fitting by polynomial does not.

\begin{figure}
    \centering
    \includegraphics[width=0.45\hsize]{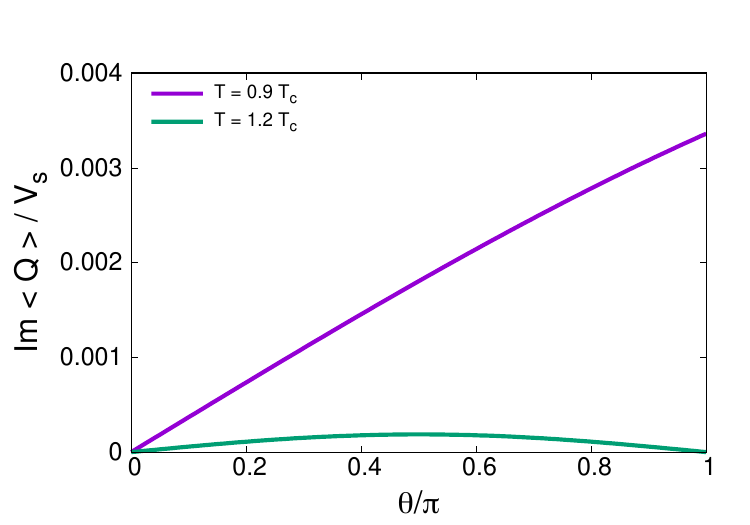}
    \caption{We perform the analytic continuation of the fitting functions.
    We choose the polynomial for $T=0.9T_{\rm c}$ case and the hyperbolic sine series for $T=1.2T_{\rm c}$, which are the better fitting functions at each temperature.}
    \label{fig:analytic_cont}
\end{figure}

Finally, we performed the analytic continuation.
In Fig.\ref{fig:analytic_cont}, we plot the analytic continuation of the fitting functions.
Here, we plot only the better fitting functions at each temperature, namely the polynomial for $T=0.9T_{\rm c}$ and the hyperbolic sine series for $T=1.2T_{\rm c}$.
Note that there is a gap at $\theta=\pi$ in the lower temperature case, but not in higher temperature case. Thus, the fact that the data points can be well fitted to the hyperbolic sine series suggests that the theory is in the CP symmetric phase at this temperature.

\begin{table}[]
    \centering
    \begin{tabular}{|c||c|c||c|} \hline
        $T$ & $a_2$ & $a_3$ & $b_3$\\ \hline \hline
        $T=0.9T_{\rm c}$ & $-1.276(58)\times 10^{-4}$ & $4.13(33) \times 10^{-6}$ & $3.29(16) \times 10^{-4}$ \\ \hline
        $T=1.2T_{\rm c}$ & $-4.6(1.0) \times 10^{-6}$ & $ 2.76(62) \times  10^{-7}$ & $1.194(75) \times 10^{-3}$ \\ \hline
    \end{tabular}
    \caption{The parameters in the fitting functions for $\Braket{Q}_{\tilde{\theta}}/V_{\rm s}$ after the infinite volume extrapolation.}
    \label{tab:fit_parm}
\end{table}

\section{Summary}
We have performed numerical studies of the 4-dimensional SU(2) Yang-Mills theory with a topological term.
In order to overcome the sign problem, we focused on the fact that the behaviour at $\theta \ne 0$ is determined in principle by the topological charge distribution at $\theta=0$.
In particular, we used the imaginary theta that enables us to study the asymptotic behaviour of the distribution.

To study the phase transition of the CP restoration, we calculated the expectation value of the topological charge $\braket{Q}_{\tilde{\theta}}$ at $\tilde{\theta} \ne 0$.
It is expected from the large-$N$ analysis that $\braket{Q}_{\tilde{\theta}} = \chi_0 \tilde{\theta}$ at zero temperature and $\braket{Q}_{\tilde{\theta}} = \chi_0 \sinh{\tilde{\theta}}$ at sufficiently high temperature.
In view of this, we fitted the extrapolated data points to the polynomial and hyperbolic sine series up to a few orders.
Our results suggest that the data points at lower temperature can be fitted by the polynomial function while those at higher temperature can be fitted by the hyperbolic sine function.
After analytic continuation, the polynomial function has a gap at $\theta=\pi$, while the hyperbolic function does not, which is consistent with the fact that the CP symmetry is restored at sufficiently high temperature.

We expect that the data points can be fitted by the polynomial function at $T\le T_{\rm CP}$ and its analytic continuation becomes zero at $\theta=\pi$ when $T=T_{\rm CP}$.
In this way, we are going to determine the critical temperature $T_{\rm CP}$ by this method.
The results will be reported in the forthcoming paper.

\acknowledgments
We would like to thank Yuta Ito for valuable discussions and comments.
The numerical calculations were carried out on Yukawa-21 at YITP in Kyoto University.
This work is also supported by the Particle, Nuclear and Astro Physics Simulation Program 
No.2021-005 (FY2021), No.2022-004 (FY2022), and No.2023-002 (FY2023) of Institute of Particle and Nuclear Studies, High Energy Accelerator Research Organization (KEK).
M.~Honda is supported by MEXT Q-LEAP, JST PRESTO Grant Number JPMJPR2117,
JSPS Grant-in-Aid for Transformative Research Areas (A) JP21H05190 and JSPS KAKENHI Grant Number 22H01222.
A.~M. is supported by JSPS Grant-in-Aid for Transformative Research Areas (A) JP21H05190.

\bibliographystyle{JHEP}
\bibliography{ref}

\end{document}